\documentclass[pra,onecolumn, showpacs, showkeys, secnumarabic, aps, amsmath, amssymb, nofootinbib, superscriptaddress, longbibliography, floatfix, table-of-contents, dblfloatfix]{revtex4-2}

\usepackage[utf8]{inputenc}
\usepackage[pdftex]{graphicx}
\usepackage{mathrsfs}
\usepackage[colorlinks, breaklinks, urlcolor={blue}, linkcolor={blue}, citecolor={blue}]{hyperref}
\usepackage{array}
\usepackage{amsmath}
\usepackage{type1cm}
\usepackage[english]{babel}
\usepackage{lmodern}
\usepackage{microtype}
\usepackage{booktabs}
\usepackage{caption}
\usepackage{braket}
\usepackage{xcolor}
\usepackage{orcidlink}
\usepackage{tikz}

\begin{document}
\title{Dephasing-Induced Distribution of Entanglement in Tripartite Quantum Systems}

\author{Sovik Roy \orcidlink{0000-0003-4334-341X} }
\email[]{s.roy2.tmsl@ticollege.org}
\affiliation{Department of Mathematics, Techno Main Salt Lake (Engg. Colg.), \\Techno India Group, EM 4/1, Sector V, Salt Lake, Kolkata  700091, India}

\author{Md. Manirul Ali\orcidlink{0000-0002-5076-7619}}
\email[]{manirul@citchennai.net}
\affiliation{Centre for Quantum Science and Technology, Chennai Institute of Technology,\\ Chennai 600069, India}

\author{Abhijit Mandal\orcidlink{0000-0001-7101-9495}}
\email[]{a.mandal1.tmsl@ticollege.org}
\affiliation{Department of Mathematics, Techno Main Salt Lake (Engg. Colg.), \\Techno India Group, EM 4/1, Sector V, Salt Lake, Kolkata  700091, India}

\author{Chandrashekar Radhakrishnan\orcidlink{0000-0001-9721-1741}}
\email[]{chandrashekar.radhakrishnan@nyu.edu}
\affiliation{Department of Computer Science and Engineering, NYU Shanghai\\, 567 West Yangsi Road, Pudong, 
Shanghai, 200126, China.}

\begin{abstract}
\noindent \textcolor{blue}{We investigate the distribution of entanglement in tripartite quantum systems under the influence of noisy environment. Entanglement distribution is an important concept to understand the robustness of entanglement in multipartite systems, while preserving multipartite entanglement amid decoherence remains a major challenge. Measuring multipartite entanglement in mixed states under decoherence is challenging. We quantify tripartite entanglement using relative entropy due to its applicability to both pure and mixed states. In this work, we explore how entanglement is distributed among the qubits of multipartite states when the three qubits are exposed to multipartite dephasing setting. In our study we use various pure and mixed tripartite states subjected to both Markovian and non-Markovian Local/Common bosonic reservoir at finite temperature. We show that the robustness of tripartite systems to decoherence depends on the distribution of entanglement and its interaction with various configurations and parameters of the bath.}
\end{abstract}


\maketitle

\section{Introduction}\label{sec:intro}
\noindent Entanglement is an incredible feature of quantum many-body systems and its
investigation holds both theoretical and practical significance \textcolor{blue}{\cite{horodeckirev2009}}.
Beyond its foundational importance, entanglement dynamics and its distribution also play an important role
in various quantum information processing tasks such as quantum teleportation
\textcolor{blue}{\cite{bennett1895}}, quantum dense coding \textcolor{blue}{\cite{bennett2881}}, quantum cryptography
\textcolor{blue}{\cite{gisin145}} and quantum computing \textcolor{blue}{\cite{nielsen2010}} and \textcolor{blue}{quantum network and internet\cite{new2023}}. The real world success
of quantum information processing, therefore, relies on the longevity of
entanglement in quantum states. However, quantum systems interact with their surroundings
environment causing decoherence, resulting to the loss of quantum features like
entanglement and coherence \textcolor{blue}{\cite{zurek715,breuer2002}}. It is therefore important to study the distribution of entanglement (which we will technically define shortly) in multipartite scenario. Various methods have
been developed to circumvent this detrimental effect of decoherence, {\it viz.}
quantum error correction \textcolor{blue}{\cite{steane1996,cory1998}}, finding decoherence-free
subspace \textcolor{blue}{\cite{zanardi1997,lidar1998}}, and dynamical decoupling \textcolor{blue}{\cite{viola1999,viola4888}}.
There is another approach which is relatively
new, known as environment engineering or reservoir engineering
\textcolor{blue}{\cite{Braun2002,Sarlette2011,Nokkala2016,mazzola79,ankim2010}}. We adopt here
this approach of engineering, the coupling between system and environment, to
enhance the longevity of multipartite entanglement. In this approach, one
can structure the reservoir spectrum in such a way that the reservoir memory
effects are significant to slow down the decay process. Our aim is to identify
various tripartite pure and mixed multipartite entangled states whose distribution of entanglement among various parties are robust against decoherence effects.\\\\
Under the influence of the environment, quantum dynamics of the system are
described by a non-unitary time evolution of the reduced density matrix, and the resulting
open system dynamics can be either Markovian or non-Markovian \textcolor{blue}{\cite{BreuerRMP16,deVega17,piilo2009,arivas2010,cdm2014,ali2014,ali2015,x1,x2,x3,x4}}.
It was shown that non-Markovian dynamics of two-qubit entangled states can be strikingly 
different from their Markovian counterparts
\textcolor{blue}{\cite{mazzola79,bellomo99,wang78,dajka2008,paz100,junli82,ali82}}. \textcolor{blue}{Extensive study of
the dynamics of entanglement in multipartite open quantum systems under the
influences of environmental effects has become matter of active research in recent
years \textcolor{blue}{\cite{ankim2010,ys2010,ma762007,aolita2008,lopez2008101,weinstein2009,christo2014,guhne2023,alim2017}}.}
One of the barriers in this type of study is the non-existence of a good measure of entanglement 
of multipartite quantum states\cite{y1,y2,y3}. A good entanglement measure should be
able to accurately reflect the operational time scale of the quantum device with
respect to a quantum information processing task. In this regard, we are going to
investigate multiparticle entanglement distribution dynamics in the presence of a structured
dephasing environment at finite temperature \textcolor{blue}{\cite{goan012111,haikka010103,gua022110}}. We have considered both the situations
when individual qubits interact with its local environment or all the qubits
collectively interact with a common environment. In this respect, the sustainability
of distribution of entanglement of various pure and mixed multipartite entangled states are examined under (a) local Markov (b) local non-Markov (c) common Markov and (d) common non-Markov dephasing
environment.\\\\
The paper is organized as follows. In section $II$, we discuss various multipartite entanglement measures and introduce the quantifier known as \textit{distribution of entanglement}. In section $III$, we describe  the Hamiltonian models that we have used to investigate three qubits subjected to local and common dephasing settings. Section $IV$ shows results and discussions on our study followed by our conclusive remarks in section $V$.
\section{Multipartite Entanglement Measure and distribution of entanglement}\label{sec:measures}
\noindent Entanglement in bipartite systems is quantified using concurrence \textcolor{blue}{\cite{wootters801998}}.  In general entanglement 
decoheres under the action of the environment and these effects can be quantified for bipartite systems using concurrence.  When investigating entanglement in multipartite systems and the effect of the environment on entanglement, we need to use more general 
multipartite measures of entanglement.  Some of the well known measures of multipartite entanglement are 
generalized concurrence \textcolor{blue}{\cite{mint260502}}, tangle \textcolor{blue}{\cite{coffman2000}}, 
$n$-tangle \textcolor{blue}{\cite{wongn044301}} , negativity \textcolor{blue}{\cite{vidal652002,leechi2003}}.  To compute the 
generalized concurrence we need to do a convex optimization.  The tangle of a tripartite system is a potential way to quantify 
the amount of tripartite entanglement in the system, but its generalization namely the $n$-tangle is not invariant under permutations 
of qubits for general odd $n$ over $3$ and hence is not a useful measure \textcolor{blue}{\cite{wongn044301}}.  On the other hand 
the negativity is a necessary and sufficient condition for two qubit conditions but for multipartite systems they are only a necessary
condition but not sufficient.  Hence these well known measures are not suitable for entanglement measurement in multipartite scenario. \\\\
An alternative approach for measuring entanglement, however, is to use the \textit{relative entropy of entanglement.}  In this method we quantify how much 
an entangled state can be operationally distinguished from the set of separable states \textcolor{blue}{\cite{vedral1997,vedral1997b}}.
This measure is intimately connected to the entanglement of distillation by providing an upper bound for it. If $\mathcal{D}$ is the set of all 
disentangled states, the measure of entanglement for a state $\rho$ is defined as 
\begin{eqnarray}
	\label{REE1}
	E(\rho) = \min_{\sigma \in \mathcal{D}}S(\rho \vert\vert \sigma),
\end{eqnarray}
where $\rho$ is the given density matrix and $\sigma$ is a separable state. The entanglement is found by measuring the 
distance of $\rho$ to the closest separable state. Here $S(\rho \vert\vert \sigma)$ is the quantum relative entropy, defined as
\begin{eqnarray}
	\label{REE2}
	S(\rho \vert\vert \sigma) = {\rm Tr} \{ \rho(\ln \rho - \ln \sigma) \},
\end{eqnarray}
and this measure works for quantum states of arbitrary dimensions.  The state $\rho$ on the boundary of separable states is called the closest 
separable state.  The measure (\ref{REE1}) gives the amount of entanglement in the density matrix from the set of disentangled states.  From 
the statistical point of view, it is known that the more entangled a state is, the more it is distinguishable from a disentangled state.\\\\  
In bipartite systems, the entire entanglement is present  between the two parties.  When we have multipartite systems, the entanglement 
can be distributed in different ways.  For example if we consider tripartite systems, the entanglement can be distributed between all 
the three qubits in a \textcolor{blue}{genuinely} multipartite fashion or in a relatively local manner.  \textcolor{blue}{In a \textcolor{blue}{genuinely} multipartite entanglement, the 
loss of a single qubit to external decoherence leads to the loss of entire entanglement in the system.  An example of such a state is 
the GHZ state where the loss of single qubit leads to a completed decohered bipartite system.  There are some multipartite systems 
in which the decoherence of a qubit does not automatically lead to the loss of entire entanglement in the system.  The $W$ state is 
a perfect example of such systems.  In these states, the three qubits are entangled in mutually pairwise fashion and so if one 
qubit decoheres, the entanglement between the other pair is still preserved.}   The entanglement distribution is a very important 
concept to understanding the robustness of entanglement in multipartite systems.  To investigate this we can  use the idea of monogamy of 
entanglement which is an inequality defined by 
\begin{equation}
    E_{A:BC} \geq E_{A:B} + E_{B:C}
\end{equation}
where $E_{A:BC}$ is the entanglement between the qubit $A$ and the bipartite block $BC$, similarly $E_{A:B}$ is the entanglement between the 
qubits $A$ and $B$ after tracing out qubit $C$.  If this inequality holds, we can observe that the genuine tripartite entanglement is 
higher, if not then the bipartite entanglements are higher.  To analyze the entanglement distribution we recast the monogamy relation to the form as shown below.
\begin{eqnarray}
\label{doe1}
    D &=& E_{A:BC} - E_{A:B} - E_{A:C},\nonumber\\
    &=& R_{A:BC} - R_{A:B} - R_{A:C},
\end{eqnarray}
where $R$ is having the expression similar to that of eq.(\ref{REE2}).
Our main objective here is to study distribution of entanglement $D$ among qubits under the impact of noisy environment. Although, $R$ itself is not monogamous, we shall continue using it for our current purpose because of its universality in applying on both pure and mixed multipartite entangled states.
We shall use the expression defined in eq.(\ref{doe1}) to investigate the distribution of entanglement in tripartite systems.  Also in our work we are studying open quantum 
systems which are subjected to dephasing environment, so the entanglement distribution will also be dynamically varying and so the 
quantity $D$ will also be a function of time.  
\section{Description of the Models}\label{sec:Model}
\noindent To study the entanglement dynamics in an open quantum system, we investigate a three qubit system in contact with an
external bath which simulates the environment.  Due to the interaction with the external environment, usually the entanglement decreases
with time.  In this work we examine the change in the distribution of entanglement $D$ of eq.(\ref{doe1}) due to the external dephasing environment.
We consider two different situations where the three qubits are in {\it (i)} local dephasing environment, i.e., each qubit in
its own dephasing environment and {\it(ii)} common dephasing environment, where all the qubits are exposed to one single dephasing
environment.

\subsection{Three Qubits Under Local Dephasing Environment}\label{subsec:Local}
\noindent We analyze a spin-boson model consisting of three non-interacting qubits, each coupled to its own local bosonic reservoir, as illustrated in 
Fig. 1. The Hamiltonian representing the three qubits and their surrounding environment is given by.
\begin{eqnarray}
\label{totalhamiltonianldp}
H = \sum_{i=1}^3\Big[\frac{\hbar}{2}\omega_{0}^i\sigma_{z}^i + \sum_{k}\hbar\omega_{ik}b_{ik}^{\dagger}b_{ik} + \sigma_{z}^i
(B_{i} + B_{i}^{\dagger})\Big].
\end{eqnarray}
Here $B_{i}=\hbar \sum_{k}g_{ik}b_{ik}$, where $\omega_{0}^i$ and $\sigma_{z}^i$ represent the transition frequency and the Pauli spin operator of the $i^{th}$ qubit, respectively. All three qubits are assumed to have the same transition frequency $\omega_{0}^{i} = \omega_{0}$. The local environment of each qubit is represented as a collection of infinite bosonic modes with frequencies $\omega_{ik}$, where $b_{ik}^{\dagger}$ and $b_{ik}$ are the creation and annihilation operators for the $k^{th}$ mode of the local environment interacting with the $i^{th}$ qubit. The coupling strength between the $i^{th}$ qubit and its local environment is denoted by $g_{ik}$. In the continuum limit, $\sum_k |g_{ik}|^2 \rightarrow \int d\omega J_{i}(\omega) \delta(\omega_{ik}-\omega)$, where $J_{i}(\omega)$ is the spectral density of the local environment for the $i^{th}$ qubit. Initially, the three-qubit system is considered decoupled from their environments, each of which starts in thermal equilibrium at temperature $T_i$. As the system evolves under the Hamiltonian $H$, the environmental degrees of freedom are traced out to obtain the reduced density matrix of the quantum system, enabling the study of its dynamics. The decay behavior of the reduced density matrix $\rho(t)$ for this three-qubit system under local dephasing is governed by the quantum master equation.
\begin{eqnarray}
\label{NL}
\frac{d}{dt} \rho(t) = \sum_{i=1}^{3} \gamma_{i}(t)  \Big( \sigma_z^i \rho(t) \sigma_z^i - \rho(t) \Big),
\end{eqnarray}
where
\begin{eqnarray}
\label{gt}
\gamma_{i}(t) = 2 \int_{0}^{\infty} d\omega  J_{i}(\omega) \coth\left(\frac{\hbar \omega}{2 k_B T_i}\right) \frac{\sin(\omega t)}{\omega}.
\end{eqnarray}
The time-dependent dephasing rate $\gamma_{i}(t)$ is influenced by the spectral density $J_{i}(\omega)$,
and we consider the following Ohmic-type spectral density \cite{leggett1987} for this model
\begin{eqnarray}
J_{i}(\omega) = \eta_{i} \omega \exp\left(-\frac{\omega}{\Lambda_i} \right).
\label{ohm}
\end{eqnarray}
It is generally assumed that the environment is sufficiently large to rapidly return to its initial state. However, we explore the impact of the reservoir's memory effect on the coherence dynamics of three qubits states with a finite cutoff frequency $\Lambda_i$. Under the Markov approximation, where the correlation time of the environment is much shorter than the timescale of the system's dynamics, the time-dependent coefficient $\gamma_{i}(t)$ can be approximated by its long-term Markov value $\gamma_{i}^M=4 \pi \eta_i k_B T_i/\hbar$. In this context, the decay dynamics of the density matrix $\rho(t)$ are considered Markovian. When uniform spectral densities and equal coupling strengths $\eta_i=\eta$ are assumed, along with equal temperatures $T_i=T$ for all local environments, the decay constant $\gamma_{i}^M$ simplifies to $\gamma_0=4 \pi \eta k_B T/\hbar$. However, the Markov assumption is valid when relaxation time of the bath is very short compared to 
system's time scale. If the bath relaxation time and the system evolution time are comparable, then the reservoir memory is not negligible
and the distribution of entanglement ($D$) of three qubits will then be governed by the master equation (\ref{NL}) with the time-dependent 
dephasing rate $\gamma_{i}(t)$ given by (\ref{gt}). 
\subsection{Three Qubits Under Common Dephasing Environment}\label{subsec:Common}
\noindent
We then explore a scenario where the three qubits interact with a shared reservoir. The microscopic Hamiltonian for the three two-level 
systems coupled to this common environment is described as
\begin{eqnarray}
\label{DephC}
H = \frac{\hbar}{2} \sum_{i=1}^{3}\Big[ \omega_0^i \sigma_z^i + \sum_k \hbar \omega_{k} b_{k}^{\dagger} 
b_{k}+ S_z \left( B + B^{\dagger} \right)\Big].
\end{eqnarray}
Here, $S_z = \sum_i \sigma_z^i$ denotes the collective spin operator for the three-qubit system, while the reservoir operator is defined as $B=\hbar \sum_k g_{k} b_{k}$. It is assumed that all three qubits share the same transition frequency, $\omega_0^i=\omega_0$. The common environment is represented as a set of bosonic field modes with frequencies $\omega_{k}$, where $b_{k}$ and $b_{k}^{\dagger}$ are the annihilation and creation operators for the $k$th mode of the environment. We begin with an initial state in which the system and environment are decoupled, and the environment is initially in thermal equilibrium at temperature $T$. The quantum master equation for the three qubits interacting with this shared environment at finite temperature is provided below
\begin{eqnarray}
\frac{d}{dt} \rho(t)\!=\!\gamma(t) S_z \rho(t) S_z\!-\!\alpha(t) S_z S_z \rho(t)\!-\!\alpha^{\ast}(t) \rho(t) S_z S_z,\nonumber\\
\label{NC}
\end{eqnarray}
where the time-dependent function $\gamma(t)$ is given by equation (\ref{gt}) and
\begin{eqnarray}
\nonumber
\alpha(t) &=& \int_{0}^{\infty} d\omega  J(\omega) \coth\left(\frac{\hbar \omega}{2 k_B T}\right) \frac{\sin(\omega t)}{\omega} \\
&-& i \int_{0}^{\infty} d\omega  J(\omega) \frac{1-\cos(\omega t)}{\omega}.
\end{eqnarray}
Under the Markov approximation, where the reservoir memory is negligible, the quantum master equation simplifies as
\begin{eqnarray}
\frac{d}{dt} \rho(t) = \frac{\gamma_0}{2}  \Big( 2 S_z \rho(t) S_z  - S_z S_z \rho(t)
-  \rho(t) S_z S_z \Big).
\label{MC}
\end{eqnarray}
The non-Markovian dynamics described in equation (\ref{NC}), which accounts for reservoir memory, differs significantly from the 
memoryless Markovian counterpart in equation (\ref{MC}). The Markovian approximation is only valid when the reservoir correlation functions decay rapidly. We then show how reservoir memory effects might improve the resilience of multiqubit coherence in a 
common environment by taking into account certain three-qubit states. We examine the coherence dynamics for an Ohmic spectral density 
$J(\omega)=\eta \omega \exp (-\omega/\Lambda)$ under both Markovian and non-Markovian dephasing. We select 
$k_B T=\hbar \omega_0/4\pi$ for our numerical data, which yields $\gamma_0=\eta \omega_0$. The other parameters are taken
as, $\eta=0.1$ and $\Lambda=10^{-2}\omega_0$ are set. Below we present a general schematic diagram of three qubits subjected to noisy environments locally and globally as well.
\begin{figure}[h]
\includegraphics[width=12.12cm]{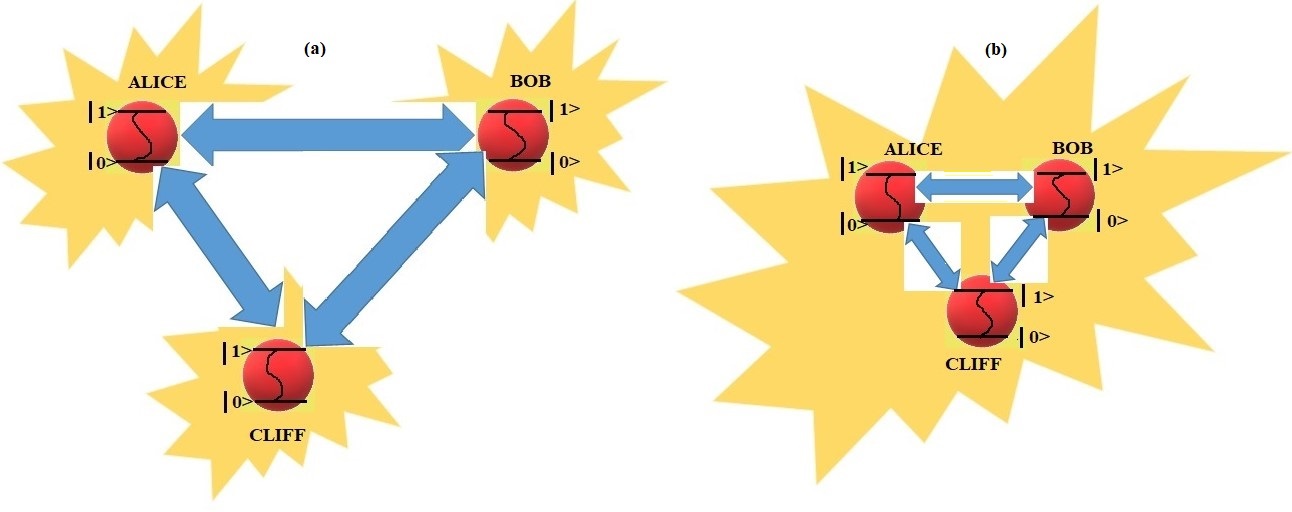}
\caption{(a) A schematic figure of tripartite states where qubits are subjected to local noise while (b) A schematic figure of tripartite states where all the three qubits are subjected to common noise.}
\label{figg}
\end{figure}
\newpage
\noindent In the left fig.$1(a)$ we see that if we consider a tripartite state, then each of its qubit is subjected to local noise while the right side of fig $1(b)$ represents the same tripartite state where all the qubits of the state are subjected to common noisy environment. The noise that has been considered in our research is the dephasing noise which shows interesting phenomena for multipartite states that are exposed to it. Another consideration that has been made here is that, these noisy environments (local/common) may be Markovian or non-Markovian. 

\section{Results and Discussion}\label{sec:results}

\subsection{Pure states under dephasing environment}
\noindent We study the distribution of entanglement $D$, defined in eq.(\ref{doe1}), for a few tripartite pure states when these states are subjected to dephasing environment. At the center of our study, we have considered tripartite pure states such as $\vert GHZ\rangle$, $\vert W\rangle$, $\vert W\bar{W}\rangle$ and $\vert Star\rangle$ \cite{ghzstate,wstate,wwbarstar}. We shall now briefly discuss the significance of these tripartite states before we proceed with our analysis.\\\\
The $\vert GHZ\rangle$ and $\vert W\rangle$ states are defined as
\begin{eqnarray}
\label{ghz}
\vert GHZ\rangle &=& \frac{1}{\sqrt{2}}\Big(\vert 000\rangle + \vert 111\rangle\Big),
\end{eqnarray} and
\begin{eqnarray}
\label{w}
\vert W\rangle &=& \frac{1}{\sqrt{3}}\Big(\vert 001\rangle + \vert 010\rangle + \vert 100\rangle\Big).
\end{eqnarray}
Based on Stochastic Local Operations and Classical Communications (SLOCC), $\vert GHZ\rangle$ and $\vert W\rangle$ are two inequivalent classes of tripartite entangled states in the sense one cannot be converted into the other \textcolor{blue}{\cite{durr2000}}. A $\vert GHZ\rangle$ state is \textcolor{blue}{genuinely} tripartite entangled state and hence when any one of its qubits is lost the resulting state completely loses its entanglement. On the other hand, in $\vert W\rangle$, genuine tripartite entanglement is missing whereas the entanglement is distributed in a bipartite way. This means that, for $\vert W\rangle$ state, when any of the qubits is subjected to decoherence, the resulting state does not loose its entanglement totally. Moreover, $\vert GHZ\rangle$ and $\vert W\rangle$ states are archetypal examples of a monogamous and polygamous state respectively. The $\vert GHZ\rangle$ state can be obtained from three polarization-entangled spatially separated photons \textcolor{blue}{\cite{dbow1999}} while the polarization entangled $\vert W\rangle$ states can be obtained by parametric down conversion from single photon source \textcolor{blue}{\cite{yt2002}}.\\\
Two other tripartite pure states of utmost significance are $\vert W\bar{W}\rangle$ and $\vert Star\rangle$ states. They are defined as
\begin{eqnarray}
\label{wwbar}
\vert W\bar{W}\rangle = \frac{1}{\sqrt{2}}\Big(\vert W\rangle + \vert \bar{W}\rangle\Big),
\end{eqnarray} and 
\begin{eqnarray}
\label{star}
\vert Star\rangle = \frac{1}{2}\Big(\vert 000\rangle + \vert 100\rangle + \vert 101\rangle + \vert 111\rangle\Big).
\end{eqnarray}
In eq.(\ref{wwbar}), $\vert \bar{W}\rangle = \frac{1}{\sqrt{3}}\Big(\vert 110\rangle + \vert 101\rangle + \vert 011\rangle\Big)$ is the spin-flipped version of $\vert W\rangle$ state. It is seen that $\vert W\bar{W}\rangle$ is the equal superposition of a standard $\vert W\rangle$ state of eqs.(\ref{w}) and its spin flipped version. The $3-$ tangle of $\vert W\bar{W}\rangle$ state is $\frac{1}{3}$ while using concurrence it can be easily shown that the entanglement of the reduced bipartite state generated from tripartite $\vert W\bar{W}\rangle$ is $\frac{1}{3}$, implying the fact that this tripartite state has both bipartite and tripartite distribution of entanglement. Now the $\vert GHZ\rangle$, $\vert W\rangle$ and $\vert W\bar{W}\rangle$ states are all symmetric states in the sense that their reduced bipartite entanglement does not depend on which of the qubits of these tripartite states has been subjected to decoherence effects. Unlike these states, $\vert Star\rangle$ state of eq.(\ref{star}) is not symmetric, since the correlations of these states are present in an asymmetric way. There are two types of qubits in $\vert Star\rangle$ state viz. (i) peripheral qubits and (ii) central qubit. The first two qubits of the state are peripheral and the third qubit is central. This means that when central qubit is traced out, the remaining qubits are left in a separable state whereas if we take partial trace over first and second qubit, entanglement is still present in the remaining qubits. The $3-$ tangle of $\vert W\bar{W}\rangle$ state is $\frac{1}{4}$ while the entanglement of the reduced bipartite state (with respect to peripheral qubits) using concurrence, is $\frac{1}{2}$ \textcolor{blue}{\cite{roy2023}}. To experimentally realize the states of eqs.(\ref{wwbar}) and (\ref{star}), polarization encoded photonic qubits are used, where horizontal and vertical polarizations are encoded as the two levels $\vert 0\rangle$ and $\vert 1\rangle$, respectively \textcolor{blue}{\cite{wwbarstar}}.
\subsection*{Distribution of entanglement of pure states:} 
\noindent The dynamics of distribution of entanglement ($D$) of the pure states described in eqs.(\ref{ghz}-\ref{star}) subjected to dephasing environment are now presented. 
\begin{figure}[h]
\includegraphics[width=11.11cm]{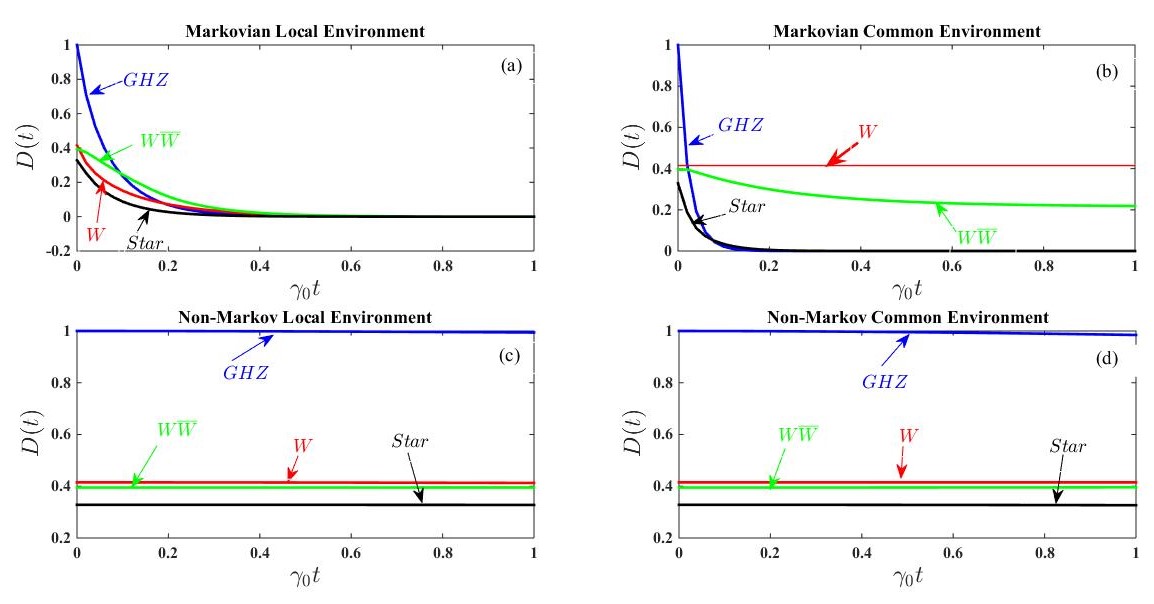}
\label{fig1}
\caption{The dynamics of distribution of entanglement of pure states such as $\vert GHZ\rangle$,$\vert W\rangle$, $\vert W\bar{W}\rangle$ and $\vert Star\rangle$ in a dephasing environment are shown for (a) local Markov,(b) common Markov, (c) local non-Markov and (d) common non-Markov environments.}
\end{figure}
\subsubsection*{Tripartite pure states under Markovian dephasing environment:}
\noindent From Fig $2(a)$, we see that, under local Markovian environment, there is decay of distribution of entanglement $D(t)$ of all pure states represented in eq.(\ref{ghz}-\ref{star}). When $\gamma_{0}t = 0$, $D(t)=1$ for $\vert GHZ\rangle$ state, $D(t)=0.4$ for both $\vert W\rangle$ and $\vert W\bar{W}\rangle$ states and that for $\vert Star\rangle$, $0.3<D(t)<0.4$. Here, the range of $\gamma_{0}t$ is considered to be $[0,1]$ and with increase in $\gamma_{0}t$, the $D(t)$ for all the pure states start decaying. The rate of decaying of $D(t)$ is much faster up to certain value of $\gamma_{0}t$ for $\vert GHZ\rangle$ state as compared to rate of decay of $D(t)$ for other three pure states. This happens until $\gamma_{0}t = 0.1$. When $0.1<\gamma_{0}t<0.4$, the $D(t)$ of $\vert W\bar{W}\rangle$ is higher than that of the remaining three pure states, while $D(t)$ of $\vert W\bar{W}\rangle$ is less than that of $\vert GHZ\rangle$ and more than both $\vert W\rangle$ and $\vert Star\rangle$ states for $0<\gamma_{0}t\leq 0.1$. Again, we see that for $0\leq \gamma_{0}t <0.4$, the $D(t)$ of $\vert W\rangle$ is higher than that of $\vert Star\rangle$ although both are decaying with same rate of decay. In the region $0\leq \gamma_{0}t<0.2$, however, $D(t)$ of $\vert GHZ\rangle$ is higher than that of $\vert W\rangle$ and $\vert Star\rangle$. The rate of decay of distribution of entanglement $D(t)$ gradually vanishes for all pure states when $\gamma_{0}t>0.5$. We can easily see that for $0.1< \gamma_{0}t\leq 0.2$, $D(\vert W\bar{W}\rangle)>D(\vert GHZ\rangle)>D(\vert W\rangle)>D(\vert Star\rangle)$ while for $0.2< \gamma_{0}t< 0.4$, $D(\vert W\bar{W}\rangle)>D(\vert W\rangle)>D(\vert GHZ\rangle)>D(\vert Star\rangle)$.\\\\
In common Markovian environment (i.e. Fig $2(b)$), the rate of decay of $D(t)$ for $\vert GHZ\rangle$ state is exponential while this rate of decay is not exponential for $\vert W\bar{W}\rangle$ state. The exponential decay rate of $\vert GHZ\rangle$ state occurs for $0<\gamma_{0}t\leq 0.1$. It is to be noted that, at $\gamma_{0}t = 0$, the $D(t)$ of $\vert GHZ\rangle$ is $1$. Also, at $\gamma_{0}t = 0$, the $D(t)$ of $\vert W\rangle$ is $0.4$ and $D(t)$ is approximately equal to $0.39$ for $\vert W\bar{W}\rangle$ state. The distribution of entanglement quantified by $D(t)$ lies between $0.2$ and $0.4$ for the $\vert Star\rangle$ state. When $0<\gamma_{0}t<0.1$, there is significant decay of $D(t)$ for the pure state $\vert Star\rangle$, while in this range, this rate of decay for $D(t)$ is not that significant for the pure state $\vert W\bar{W}\rangle$. However it is very interesting to note that for $0\leq \gamma_{0}t \leq 1$, there is no decay in the distribution of entanglement $D(t)$ for $\vert W\rangle$ state. We know $\vert W\bar{W}\rangle$ is the linear superposition of $\vert W\rangle$ and spin flipped version $\vert \bar{W}\rangle$, but the decay of $D(t)$ are not same for the two pure states. For pure state $\vert W\rangle$ undergoing through dephasing channel, there is no decay in $D(t)$, but this is not the case for $\vert W\bar{W}\rangle$ state. We see that, there is significant decay of $D(t)$ for $\vert W\bar{W}\rangle$ as compared to $\vert W\rangle$. For $\gamma_{0}t>0.7$, the decay of all the pure states subjected to the dephasing environment, reach equilibrium. This is also to be noted that for $\gamma_{0}t\geq 0.2$, the $D(t)$ of $\vert GHZ\rangle$ and $\vert Star\rangle$ states decays to zero, while that of $\vert W\rangle$ and $\vert \bar{W}\rangle$ do not.
\subsubsection*{Tripartite pure states under non-Markovian dephasing environment:}
\noindent For the pure states in local non-Markovian dephasing environment (Fig.$2(c)$) we observe that there is no decay of $D(t)$ for $0\leq \gamma_{0}t \leq 1$. However, from Fig.$2(d)$ (i.e. common non-Markov dephasing environment), in the case of three pure states viz. $\vert W\rangle$, $\vert W\bar{W}\rangle$ and $\vert Star\rangle$, there are no decays in the distribution of entanglement $D(t)$. Moreover, though there is almost no decay of $D(t)$ in case of $\vert GHZ\rangle$ state, decaying is observed for this pure state when $\gamma_{0}t>0.6$. For both local and common non-Markovian dephasing environments thus we see that, all three pure states behave similarly except for $\vert GHZ\rangle$ state in the common non-Markovian environment for $\gamma_{0}t>0.6$. The amount of distribution of entanglement $D(t)$, measured under local and common non-Markovian regime, for the entire range of $\gamma_{0}t$, can be given through the ordering relation $D(\vert GHZ\rangle)>D(\vert W\rangle)>D(\vert W\bar{W}\rangle)>D(\vert Star\rangle)$.
\subsection{Mixed states under dephasing environment}
\noindent Sustainability of pure states are hard to be achieved from experimental view point. Keeping this in mind we have carried forward with our analysis on dynamics of distribution of entanglement, which has so far been studied in the case of pure tripartite states, to some mixed states that have been designed as mixtures of well-defined and significant tripartite states. 
In this regard, we have taken into consideration a few mixed states namely (i) mixture of $\vert GHZ\rangle$ and $\vert W\rangle$
state, (ii) mixture of Werner and $\vert GHZ\rangle$ and (iii) mixture of Werner and $\vert W\rangle$ state \textcolor{blue}{\cite{junge024306,lohm260502,2104arxiv}}. These states are respectively shown below. 
\begin{eqnarray}
\label{ghz-werner}
\rho^{wer}_{ghz} = p\vert GHZ\rangle\langle GHZ\vert + \frac{1-p}{8}\mathbb{I}_{8}, 
\end{eqnarray}
\begin{eqnarray}
\label{w-werner}
\rho^{wer}_{w} = p\vert W\rangle\langle W\vert + \frac{1-p}{8}\mathbb{I}_{8}, 
\end{eqnarray}
and
\begin{eqnarray}
\label{ghz-w}
\rho^{ghz}_{w} = p\vert GHZ\rangle\langle GHZ\vert + (1-p)\vert W\rangle\langle W\vert.
\end{eqnarray}
Here, $p$ is the probability of mixing and $\mathbb{I}_{8}$ is the $8\times 8$ identity matrix. The states described in eqs.(\ref{ghz-werner}) and (\ref{w-werner}) are sometimes known as generalized Werner states. The mixtures of $\vert GHZ\rangle$ and $\vert W\rangle$  with completely unpolarized state (or maximally mixed state or sometimes referred to as white noise) have been taken respectively. Thus the two mixed states of eqs.(\ref{ghz-werner}) and (\ref{w-werner}) are mixtures of regular tripartite states with white noise. The third state described in eq.(\ref{ghz-w}) is a statistical mixture of $\vert GHZ\rangle$ and $\vert W\rangle$ states. 

\subsection*{Distribution of entanglement of mixed states:} 
\noindent The dynamics of distribution of entanglement of the mixed states described in eqs.(\ref{ghz-werner}-\ref{ghz-w}) subjected to dephasing environment are now presented. 
\subsubsection*{Tripartite mixed states $\rho^{wer}_{ghz}$ under Markovian dephasing}
\begin{figure}[h]
\includegraphics[width=11.11cm]{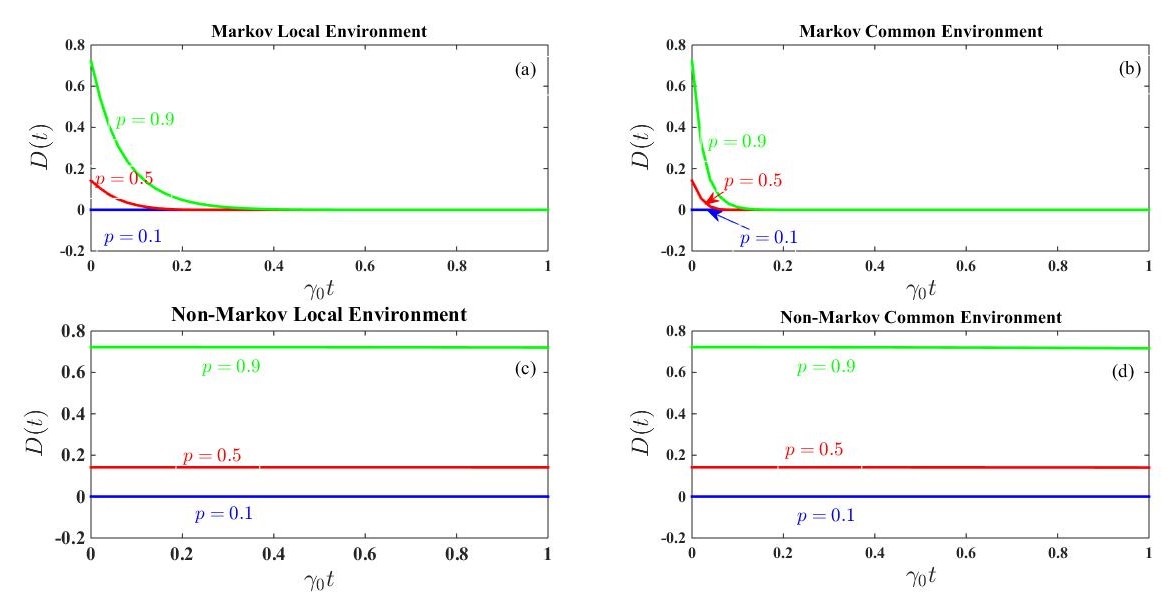}
\caption{The dynamics of distribution of entanglement of mixed state $\rho^{wer}_{ghz}$ in a dephasing environment are shown for (a) local Markov,(b) common Markov, (c) local non-Markov and (d) common non-Markov environments.}
\label{fig2}
\end{figure}
\noindent Fig.$3(a)$ depicts the nature of decaying of distribution of entanglement of mixed state $\rho^{wer}_{ghz}$ of eq.(\ref{ghz-werner}). From fig.$3(a)$ (which represents the local Markov dephasing environment) we see that for mixing parameter $p=0.9$ the $D(t)$ of $\rho^{wer}_{ghz}$ decays as $\gamma_{0}t$ increases. At $\gamma_{0}t = 0$, $D(\rho^{wer}_{ghz}) = 0.7$, then the decaying process starts and ultimately $D(t)$ vanishes when $\gamma_{0} t > 0.3$.  Likewise for mixing parameter $p = 0.5$, $D(t)$ starts as $0.18$ and then decreases to zero as $\gamma_{0}t$ increases. But for mixing parameter $p = 0.1$, the distribution of entanglement $D(t)$ for $\rho^{wer}_{ghz}$ is always zero. It is also to be noted that $D(\rho^{wer}_{ghz})_{p = 0.9}> D$ ($\rho^{wer}_{ghz})_{p = 0.5}$ for $0<\gamma_{0}t < 0.3$.\\\\
Also from Fig.$3(b)$ we see that for $\rho^{wer}_{ghz}$ subjected to common dephasing environment $D(\rho^{wer}_{ghz})_{p=0.9}>D(\rho^{wer}_{ghz})_{p=0.5}$ when $0\leq \gamma_{0}t\leq 0.1$. The distribution of entanglement starts at $0.7$ for mixing parameter $p = 0.9$ and at $0.18$ for mixing parameter $p = 0.5$ at $\gamma_{0}t = 0$ and then start decaying as $\gamma_{0}t$ increases. However, just like local Markov environment, in common Markovian environment too, for mixing parameter $p = 0.1$, distribution of entanglement $D(t)$ of the mixed state $\rho^{wer}_{ghz}$ is zero throughout the domain $[0,1]$ of $\gamma_{0}t$.\\\\
The  amount of distribution of entanglement of mixed state $\rho^{wer}_{ghz}$, with variation in $\gamma_{0}t$ for both local and common Markovian dephasing environment, decays but the rate of decay in local Markovian bath is slower than the rate of decay in common Markovian bath.
\subsubsection*{Tripartite mixed states $\rho^{wer}_{ghz}$ under Non-Markovian dephasing}
\noindent We see from Figs. $3(c)$ and $3(d)$ that, in both the non-Markovian local and common dephasing environments, for mixing parameter $p = 0.9$ and for $p = 0.5$, the value of the distribution of entanglement of the state $\rho^{wer}_{ghz}$ are respectively $0.7$ and $0.19$. Consequently, there is no decay of $D(t)$ as $\gamma_{0}t$ increases. However, for mixing parameter $p=0.1$, the distribution of entanglement $D(t)$ of the given mixed state is always zero, when $0\leq \gamma_{0}t\leq 1$.
\subsubsection*{Tripartite mixed states $\rho^{wer}_{w}$ under Markovian and Non-Markovian dephasing}
\begin{figure}[h]
\begin{center}
\includegraphics[width=11.11cm]{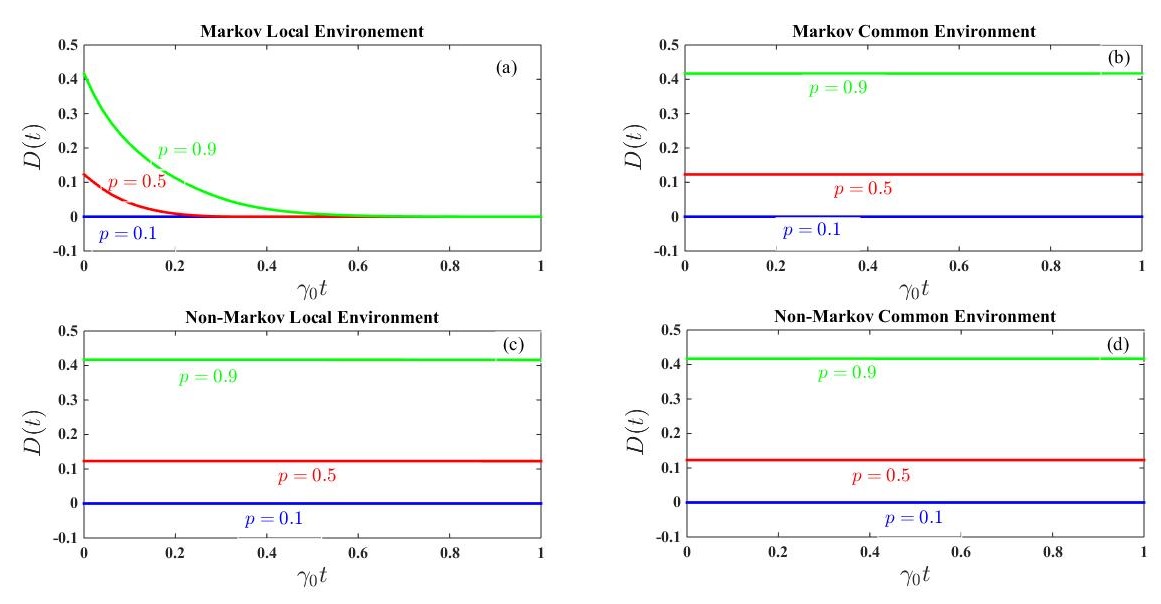}
\end{center}
\caption{The dynamics of distribution of entanglement of mixed state $\rho^{wer}_{w}$ in a dephasing environment are shown for (a) local Markov,(b) common Markov, (c) local non-Markov and (d) common non-Markov environments.}
\label{fig3}
\end{figure}
\noindent The case of mixed state $\rho^{wer}_{w}$ subjected to dephasing environment is rather interesting. We see that in common Markovian environment as well as in both local and common non-Markovian environment, there are no decaying of $D(t)$ with $\gamma_{0}t>0$. In all these three cases, $D(t)$ takes the value of $0.42$ and $0.12$ [as shown in figs. $4(b),\: 4(c)$ and $4(d)$] for mixing parameters $p = 0.9$ and $p = 0.5$. Again, in all these three environments for mixing parameter $p = 0.1$, the distribution of entanglement $D(t) = 0$ when $0\leq \gamma_{0}t\leq 1$. However, in the case of local Markovian dephasing environment only (fig. $3(a)$), we see that when $0\leq \gamma_{0}t\leq 0.6$, $D(\rho^{wer}_{w})_{p=0.9}>D(\rho^{wer}_{w})_{p=0.5}$ and as $\gamma_{0}t> 0$, we observe decaying process in $D(t)$ for both $p = 0.9$ and $p = 0.5$. Ultimately $D(t)$ vanishes with $\gamma_{0}t > 0.5$ for mixing parameter $p = 0.9$ and with $\gamma_{0}t > 0.2$ for mixing parameter $p = 0.5$. However, for mixing parameter, $p = 0.1$, $D(t)$ is always zero when the mixed states $\rho^{wer}_{w}$ is exposed to local Markovian dephasing environment.
\subsubsection*{Tripartite mixed states $\rho^{ghz}_{w}$ under Markovian dephasing}
\begin{figure}[h]
\begin{center}
\includegraphics[width=11.11cm]{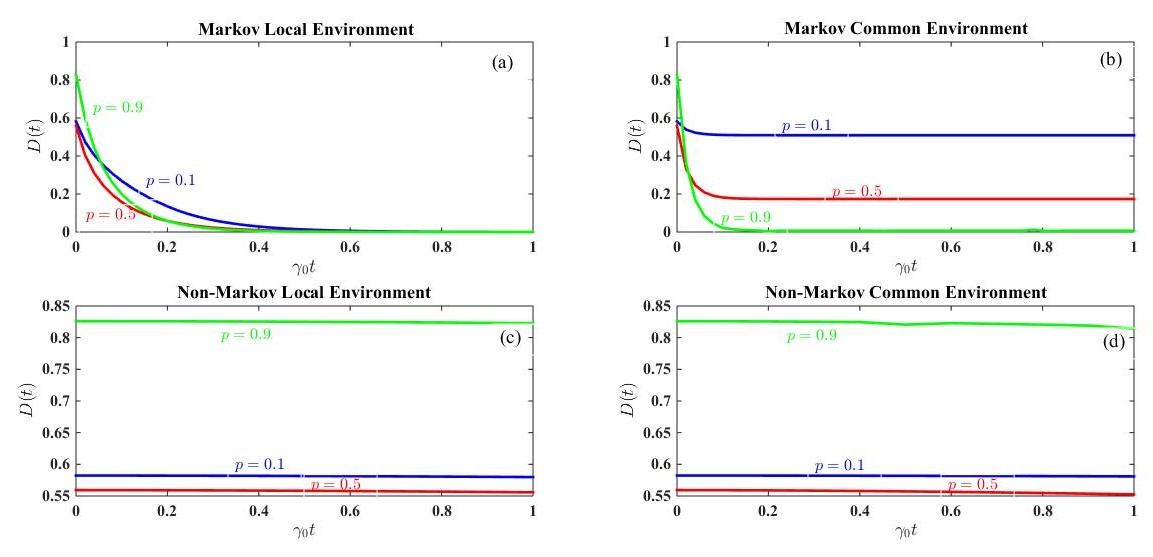}
\end{center}
\caption{The dynamics of distribution of entanglement of mixed state $\rho^{ghz}_{w}$ in a dephasing environment are shown for (a) local Markov,(b) common Markov, (c) local non-Markov and (d) common non-Markov environments.}
\label{fig4}
\end{figure}
\noindent Fig.$5(a)$ depicts local Markovian dephasing environment and fig.$5(b)$ represents common Markovian dephasing environment for mixed state $\rho^{ghz}_{w}$. We see from fig.$5(a)$ that at $\gamma_{0}t = 0$, $D(\rho^{ghz}_{w})_{p=0.9}>D(\rho^{ghz}_{w})_{p=0.1}>D(\rho^{ghz}_{w})_{p=0.5}$. The decaying process starts then. For the range $0<\gamma_{0}t<0.2$, the decaying is faster for $p = 0.9$. In the range $0.1<\gamma_{0}t<0.5$, however, $D(\rho^{ghz}_{w})_{p=0.1}>D(\rho^{ghz}_{w})_{p=0.9}$. When $\gamma_{0}t \geq 0.2$, $D(\rho^{ghz}_{w})_{p=0.9} = D(\rho^{ghz}_{w})_{p=0.5}$ and these two ultimately vanish when $\gamma_{0}t\geq 0.4$. Moreover, for mixing parameter $p=0.1$, the decaying of distribution of entanglement vanishes when $\gamma_{0}t>0.5$.
Fig.$4(b)$ shows that, at $\gamma_{0}t =0$, $D(\rho^{ghz}_{w})_{p=0.9}>D(\rho^{ghz}_{w})_{p=0.1}>D((\rho^{ghz}_{w})_{p=0.5}$. The decaying of $D(t)$ is exponential when $0<\gamma_{0}t<0.1$ and vanishes  at $\gamma_{0}t = 0.2$ for mixing parameter $p=0.9$. But it is interesting to note that, although a slight decay is observed for mixing parameter $p=0.1$ when $0<\gamma_{0}t<0.1$, after that there is no decay and distribution of entanglement attains a fixed value throughout $\gamma_{0}t>0.1$. Likewise, we see that the rate of decay of $D(t)$ for mixing parameter $p=0.5$ is higher than that for $p=0.1$ when $0<\gamma_{0}t<0.1$. But for $\gamma_{0}t>0.1$, there is no further decay and $D(t)$ attains a fixed value throughout the range where $\gamma_{0}t>0.1$.
\subsubsection*{Tripartite mixed states $\rho^{ghz}_{w}$ under Non-Markovian dephasing}
\noindent From Figs.$5(c)$ and $5(d)$, we see that both in local non-Markovian environment and in common non-Markovian environment, at $\gamma_{0}t =0$, the $D(t)$ for $\rho^{ghz}_{w}$ lies in the following range. For mixing parameter $p=0.9$, $0.8<D(t)<0.85$ and for mixing parameter $p = 0.5$, $D(t) = 0.55$ while for $p=0.1$, $0.55<D(t)<0.6$, that is $D(\rho^{ghz}_{w})_{p=0.9}>D(\rho^{ghz}_{w})_{p=0.1}>D(\rho^{ghz}_{w})_{p=0.5}$. In local non-Markov environment, there is no decay in $D(t)$. But in common non-Markov environment for $p = 0.9$, there is slight decaying as $\gamma_{0}t>0.4$ and for $p=0.5$, decay occurs when $\gamma_{0}t>0.7$ (which is clear from fig.$5(d)$).
\section{Conclusion:}
\noindent In quantum information and computation, it is crucial to study the entanglement of the quantum states, as entanglement is the key feature which is responsible for various quantum information processing tasks. When these quantum states are subjected to realistic environment the effects on the entanglement of these quantum states are, therefore, matter of concerns. In multipartite system (in our case it is tripartite), we often are very interested in studying the distribution of entanglement over various bipartite cuts. It is because, in multipartite scenario, dynamical stability of distribution of entanglement is crucial. We have defined here a quantity known as distribution of entanglement,$D$, based on the measure of relative entropy of entanglement. Relative entropy of entanglement has been considered here, since this measure can be used for both pure and mixed tripartite states. An important question thus arises regarding the preservation of distribution of entanglement in a given tripartite system. Hence we have carefully chosen some specific tripartite pure and mixed states initially and also a specific environmental condition, known as dephasing environment to conduct our research. In this study, we examine sustainability of distribution of entanglement in a tripartite system subjected to an external dephasing environment. This dephasing environment, which arises from phase damping processes, is critical in describing decoherence in realistic systems with significant practical implications. The dynamics of distribution of entanglement of these different tripartite pure and mixed states are observed under dephasing environment.  Additionally, we consider two distinct scenarios, (a) where each qubit is immersed in its own individual noisy environment, and (b) all qubits share a common noisy environment. The entanglement dynamics are influenced by the nature of the environment, specifically whether it is Markovian or non-Markovian.  
\\\\
Our analysis of the distribution of entanglement in tripartite pure states, subjected to the (local/common) Markovian/non-Markovian environments under the effects of dephasing environment, are as follows. We see that, pure states such as $\vert GHZ\rangle$, $W\bar{W}$ and $\vert Star\rangle$ states exhibit decaying of the distribution of entanglement with variation in time in memoryless Markovian environment. However, the dephasing environment induces decay in the distribution of entanglement in $\vert W\rangle$, when subjected to local Markovian environment. On the other hand, in common Markovian environment such decaying effects are absent for $\vert W\rangle$. It is interesting to note down that $\vert W\rangle$ and $\vert W\bar{W}\rangle$ are behaving differently here. We know that $\vert W\bar{W}\rangle$ is the linear superposition of $\vert W\rangle$ and its spin flipped version $\vert \bar{W}\rangle$, but the two states are behaving in a different manner with respect to distribution of entanglement under dephasing effect in common Markovian bath. In local non-Markovian environment dephasing cannot induce any decay effect while in common Markovian environment decay of distribution of entanglement is observed for $\vert GHZ\rangle$ only when $\gamma_{0}(t)$ exceeds $0.8$. Now we give our conclusion on decaying effect of dephasing environment on mixtures of Werner-GHZ ($\rho^{wer}_{ghz}$), Werner-W ($\rho^{wer}_{w}$) and GHZ-W ($\rho^{ghz}_{w}$). We observe that in non-Markovian local/common bath there are no decaying effects of distribution of entanglement on the mixtures $\rho^{wer}_{ghz}$ and $\rho^{wer}_{w}$. But in the same non-Markovian common environment there is decay induced by dephasing environment on the mixture $\rho^{ghz}_{w}$. In the local Markov environment, we see that there is decay of $D(t)$ for mixing parameters $p = 0.9$ and $p = 0.5$ for the state $\rho^{wer}_{ghz}$ and $\rho^{wer}_{w}$ (w.r.t both local/common environment). It is interesting to note down that when $\rho^{wer}_{w}$ is subjected to common Markovian bath, $\rho^{wer}_{w}$ experiences no decay in $D(t)$ [see fig. $4(b)$], whereas for other mixed states subjected to common Markovian environment exhibit decay of distribution of entanglement over time. For the mixed state $\rho^{ghz}_{w}$, there is decay induced by dephasing environment when the given state is subjected to common Markovian bath but the decay is not sharp and stabilizes itself at some value when $\gamma_{0}(t)$ exceeds certain range. This happens when the values of mixing parameters $p = 0.9$ and $p = 0.5$. Our study unlocks the possibilities of exploring the tripartite pure and mixed states being subjected to Markovian and non-Markovian bath from experimental view points as these states are viable candidates for processing quantum information. In future one can use squashed entanglement which is monogamous and continue with the similar kind of study applicable on tripartite states.\\\\
{\it Data availability statement}: All data that support the findings of this study are included within the article (and
any supplementary files).


\begin{thebibliography}{99}
\bibitem{horodeckirev2009} Horodecki, R., Horodecki, P., Horodecki, M. and Horodecki K., \textit{Rev. of Mod. Phys.}, 81, 865, (2009).
\bibitem{bennett1895} Bennett, C.H., Brassard, G., Cr\'{e}peau, C., Jozsa, R., Peres, A and Wootters, W. K., \textit{Phys. Rev. Lett.}, 70, 1895, (1993).
\bibitem{bennett2881} Bennett, C.H. and Wiesner, S.J., \textit{Phys. Rev. Lett.}, 69, 2881, (1992).
\bibitem{gisin145} Gisin, N., Ribordy, G., Tittel, W. and Zbinden, H., \textit{Rev. Mod. Phys.} 74, 145, (2002).
\bibitem{nielsen2010} Nielsen, M. A. and Chuang, I. L., \textit{Quantum computation and quantum information}, Cambridge University Press, (2010)
\textcolor{blue}{\bibitem{new2023} Azuma, K., Economou, S.E., Elkouss, D., Hilaire, P., Jiang, L., and Kwong, H., \textit{Rev. Mod. Phys.}, 95, 045006, (2023).}
\bibitem{zurek715} Zurek, W., \textit{Rev. Mod. Phys.}, 75, 715, (2003).
\bibitem{breuer2002} Breuer H.-P. and Petruccione F., Theory of Open Quantum Systems, \textit{Oxford University Press}, 2002.
\bibitem{steane1996} Steane, A. M., \textit{Phys. Rev. Lett.}, 77, 793, (1996).
\bibitem{cory1998} Cory, D. G., Price, M., Maas, W., Knill, E., Laflamme, R., Zurek, W. H., Havel, T. F., and Somaroo, S. S., \textit{Phys. Rev. Lett.}, 81, 2152, (1998).
\bibitem{zanardi1997} Zanardi, P., and Rasetti, M., \textit{Phys. Rev. Lett.}, 79, 3306, (1997).
\bibitem{lidar1998} Lidar, D. A., Chuang, I. L., and Whaley, K. B., \textit{Phys. Rev. Lett.}, 81, 2594, (1998).
\bibitem{viola1999} Viola, L., Knill, E., and Lloyd, S., \textit{Phys. Rev. Lett.}, 82, 2417, (1999).
\bibitem{viola4888} Viola, L., Llyod, S. and Knill, E., \textit{Phys. Rev. Lett.}, 83, 4888, (1999).
\bibitem{Braun2002} Braun, D., \textit{Phys. Rev. Lett.}, 89, 277901, (2002)
\bibitem{Sarlette2011} Sarlette, A., Raimon, J-M., Brune, M. and Rouchon, P., \textit{Phys. Rev. Lett.},  107, 010402, (2011).
\bibitem{Nokkala2016} Nokkala, J., Galve, F., Zambrini, R., Maniscalco, S., and Pillo, J., \textit{Sci. Rep.}, 6. 26861, (2016).
\bibitem{mazzola79} Mazzola, L., Maniscalco, S., Pillo, J., Suominen, K. A., and Garraway, B. M., \textit{Phys. Rev. A}, 79, 042302, (2009).
\bibitem{ankim2010} An, N. B., Kim, J., and Kim, K., \textit{Phys. Rev. A}, 82, 032316, (2002).
\bibitem{BreuerRMP16} Breuer, H.- P, Laine, E.-M, Pillo, J., and Vacchini, B., \textit{Rev. Mod. Phys.}, 88, 021002, (2002).
\bibitem{deVega17} De. Vega, I., and Alonso, D., \textit{Rev. Mod. Phys.}, 89, 015001, (2017).
\bibitem{piilo2009} Breuer, H-P., Laine, E-M. and Piilo, J., \textit{Phys. Rev. Lett.}, 103, 210401, (2009).
\bibitem{arivas2010} De. Vega, I., and Alonso, D., \textit{Phys. Rev. Lett.}, 105, 050403, (2010).
\bibitem{cdm2014} De. Vega, I., and Alonso, D., \textit{Phys. Rev. Lett.}, 112, 120404, (2014).
\textcolor{blue}{\bibitem{ali2014} Chen, P-O., and Ali. M-M, \textit{Sci. Rep.}, 4, 6165, (2014).}
\textcolor{blue}{\bibitem{ali2015} Ali, M-M., Lo, P-Y., Tu, M. W-Y.,, and Zhang, W-M., \textit{Phys. Rev. A}, 92, 062306, (2015).}
\textcolor{blue}{\bibitem{x1} Hou, S.C., Liang, S.L., and Yi, X-X., \textit{Phys. Rev. A}, 92, 012109, (2015).}
\textcolor{blue}{\bibitem{x2} Wang, D., Huang, A-J., Hoehn, R.D., Ming, F., Sun, W-Y., Shi, J-D., Ye, L., and Kais, S., \textit{Sci. Rep.}, 7, 1066, (2017).}
\textcolor{blue}{\bibitem{x3} Wang, D., Shi, W-N., Hoehn R.D., Ming, F., Sun, W-Y., Ye, L. and Kais, S., \textit{Quant. Inf. Proc.}, 17, 335, (2018).}
\textcolor{blue}{\bibitem{x4} Chen, M-N., Wang, D. and Ye, L., \textit{Phys. Lett. A}, 383, 10, (2019).
\bibitem{bellomo99} Bellomo, B., Franco, R. L., and Compagno, G., \textit{Phys. Rev. Lett.}, 99, 160502, (2007).}
\bibitem{wang78} Wang, F.-Q, Zhang, Z.-M, and Liang, R.-S., \textit{Phys. Rev. A}, 78, 062318, (2007).
\bibitem{dajka2008} Dajka, J., Mierzejewski, M., and Luczka, R.-S., \textit{Phys. Rev. A}, 77, 042316, (2008).
\bibitem{paz100} Paz, J. P., and Roncaglia, A. J., \textit{Phys. Rev. Lett.}, 100, 220401, (2008).
\bibitem{junli82} Li, J.-G, Zou, J. , and Shao, B., \textit{Phys. Rev. A}, 82, 042318, (2002).
\bibitem{ali82} Ali, M.-M, Chen, P.-W, and Goan, H.-S., \textit{Phys. Rev. A}, 82, , 022103, (2010).
\bibitem{ys2010} Weinstein, Y. S., \textit{Phys. Rev. A}, 82, 032326, (2010).
\bibitem{ma762007} San Ma, X., Wang, A. M., and Cao, Y., \textit{Phys. Rev. B}, 76, 155327, (2007).
\bibitem{aolita2008} Aolita, L., Chaves, R., Cavalcanti, D., Acin, A., and Daviddovich, L., \textit{Phys. Rev. Lett.}, 100, 080501, (2008).
\bibitem{lopez2008101} Lopez, C., Romero, G., Lastra, F., Solanko, E., and Re-Tamal, J. \textit{Phys. Rev. Lett.}, 101, 080503, (2008).
\bibitem{weinstein2009} Weinstein, Y. S., \textit{Phys. Rev. A}, 79, 012318, (2009).
\bibitem{christo2014} Eltschka, C., Braun, D., and Siewert, J., \textit{Phys. Rev. A}, 89, 062307, (2014).
\textcolor{blue}{\bibitem{guhne2023} Wyderka, N., Ketterer, A., Imai, S., Bonsel, J.L., Jones, D. E., Kirby, B.T., Yu, X-D., and Guhne, O, \textit{Phys. Rev. Lett.}, 131, 090201, (2023).}
\textcolor{blue}{\bibitem{alim2017} Ali, M., \textit{Int. Jour. Quant. Inf.}, 15 (3), 1750022, (2017).}
\textcolor{blue}{\bibitem{y1} Dong, D.D., Song, X-K,  Fan, X.G.,  Ye, L., and Wang, D., \textit{Phys. Rev. A}, 107, 052403, (2023).}
\textcolor{blue}{\bibitem{y2} Dong, D.D., Li, L-J., Song, X-K., Ye, L., and Wang, D., \textit{Phys. Rev. A}, 110, 032420, (2024).}
\textcolor{blue}{\bibitem{y3} Li, L-J., Fan, X.G., Song, X-K,  Ye, L., and Wang, D., \textit{Phys. Rev. A}, 110, 012418, (2024).}
\bibitem{goan012111} Goan, H.-S., Jian, C.-C., and Chen, P.-W, \textit{Phys. Rev. A}, 82, 012111, 2010.
\bibitem{haikka010103} Haikka, P., Johnson, T., and Maniscalco, S., \textit{Phys. Rev. A}, 87, 010103, (2013).
\bibitem{gua022110} Guarnieri, G., Smirne, A., and Vacchini, B., \textit{Phys. Rev. A}, 90, 022110, (2014).
\bibitem{wootters801998} Wootters, W. K., \textit{Phys. Rev. Lett.}, 80, 2245.
\bibitem{mint260502} Mintert, F., Kus, M., and Buchleitner, \textit{Phys. Rev. Lett.}, 95, 260502, (2005).
\bibitem{coffman2000} Coffman, V., Kundu, J., and Wootters, W. K., \textit{Phys. Rev. A}, 61, 052306, (2000).
\bibitem{wongn044301} Wong, A., and Christensen, N., \textit{Phys. Rev. A}, 63, 044301, (2001).
\bibitem{vidal652002} Vidal, G., and Werner, R. F., \textit{Phys. Rev. A}, 65, 032314, (2002).
\bibitem{leechi2003} Lee, S., Chi, D.P., Oh, S.D., and Kim, J., \textit{Phys. Rev. A}, 68, 062304, (2003).
\bibitem{vedral1997} Vedral, V., Plenio, M. B., Rippin, M. A.and Knight, P. L., \textit{Phys. Rev. Lett.}, 78, 2275, (1997).
\bibitem{vedral1997b} Vedral, V., and Plenio, M. B., \textit{Phys. Rev. A}, 57, 1619, (1998).
\bibitem{vedral2002} Vedral, V., \textit{Rev. Mod. Phys.}, 74, 1, (2002).
\bibitem{leggett1987} Leggett, A.J., Chakravarty, S., Dorsey, A. T., Fisher, M.P., Garg, A., and Zwerger, W., \textit{Rev. of Mod. Phys.}, 59,, 1, (1987). 
\bibitem{ghzstate} Greenberger, D. M., Horne, M. A., and Zeilinger, A.,  \textit{Bell's theorem, Quantum theory and conceptions of the universe}, ed. M. Kafatos (Kluwer, Dordrecht, p.69), (1989).
\bibitem{wstate} D$\ddot{u}$r, W., and Cirac, J.I., \textit{Phys. Rev. A}, 61, 042314, (2000).
\bibitem{wwbarstar} Cao, H., Radhakrishnan, C., Ming, S., Ali, M.-M., Zhang, C., Huang, Y.-F, Byrnes, T., Li, C.-F., and Guo, G.-C., \textit{Phys. Rev. A}, 102, 012403, (2020).
\bibitem{durr2000} D$\ddot{u}$r, W., Vidal, G., and Cirac, J. I., \textit{Phys. Rev. A}, 62, 062314, (2000).
\bibitem{dbow1999} Bouwmeester, D., Pan, J.-W., Daniell, M., Weinfurter, H., and Zeilinger, A., \textit{Phys. Rev. Lett.}, 82, 1345, (1999).
\bibitem{yt2002} Yamamoto, T., Tamaki, K., Koashi, M., and Imoto, N., \textit{Phys. Rev. A}, 66, 064301, (2002).
\bibitem{roy2023} Roy, S., Bhattacharjee, A., Radhakrishnan, C., Ali, M.-M., and Ghosh, B., \textit{Int. Jour. Quant. Inf.}, 21, 2, (2023).
\bibitem{junge024306} Jung, E., Hwang, M.-R, Park, D., and Son, J.-W., \textit{Phys. Rev. A}, 79, 024306, (2009).
\bibitem{lohm260502} Lohmayer, R., Osterloh, A., Siewart, J., and Uhlmann, A., \textit{Phys. Rev. Lett.}, 97, 260502, (2006).
\bibitem{2104arxiv} Czerwinski, A., \textit{Int. Jour. Mod. Phys. B}, 36, 19, (2022).
\end{thebibliography}
\end{document}